\documentclass[manuscript,screen]{acmart}
\graphicspath{{./fig/}}
\usepackage{soul,color}
\definecolor{redd}{rgb}{0.858, 0.188, 0.478}
\usepackage{tcolorbox}
\tcbuselibrary{listings,breakable}
\usepackage{algorithm,algorithmic,amsmath}

\newtcolorbox{highlighted}{colback=yellow,breakable}

\newcommand{\tool}{\emph{SPI}}
\newcommand{\jingkai}[1]{#1}
\newcommand{\jingkaicyan}[1]{#1}
\newcommand{\jingkaiorange}[1]{#1}

\AtBeginDocument{%
  \providecommand\BibTeX{{%
    \normalfont B\kern-0.5em{\scshape i\kern-0.25em b}\kern-0.8em\TeX}}}

\setcopyright{acmcopyright}
\acmJournal{TOSEM}
\acmYear{2021} \acmVolume{1} \acmNumber{1} \acmArticle{1} \acmMonth{1} \acmPrice{15.00}\acmDOI{10.1145/3468854}

\begin{document}

\title{SPI: Automated Identification of Security Patches via Commits}

\author{Yaqin Zhou}
\authornote{indicate equal contribution}
\email{yaqinchou@gmail.com}
\affiliation{%
  \institution{Nanyang Technological University}
  \country{Singapore}
}

\author{Jing Kai Siow}
\authornotemark[1]
\affiliation{%
  \institution{Nanyang Technological University}
  \country{Singapore}}
\email{jingkai001@e.ntu.edu.sg}

\author{Chenyu Wang}
\affiliation{
  \institution{Nanyang Technological University}
  \country{Singapore}}
\email{CWANG014@e.ntu.edu.sg}

\author{Shangqing Liu}
\affiliation{%
  \institution{Nanyang Technological University}
  \country{Singapore}}
\email{shangqin001@e.ntu.edu.sg}

\author{Yang Liu}
\affiliation{%
  \institution{Nanyang Technological University}
  \country{Singapore}}
\email{yangliu@ntu.edu.sg}


\begin{abstract}
Security patches in open-source software, providing security fixes to identified vulnerabilities, are crucial in protecting against cyber attacks. Security advisories and announcements are often publicly released to inform the users about \jingkaiorange{potential security vulnerability.} Despite the National Vulnerability Database (NVD) publishes identified vulnerabilities, a vast majority of vulnerabilities and their corresponding security patches remain beyond public exposure, e.g., in the open-source libraries that are heavily relied on by developers. \jingkaicyan{As many of these patches exist in open-sourced projects, the problem of curating and gathering security patches can be difficult due to their hidden nature. An extensive and complete security patches dataset could help end-users such as security companies, \textit{e.g.,} building a security knowledge base, or researcher, \textit{e.g.,} aiding in vulnerability research.}

To efficiently curate security patches including \jingkai{undisclosed patches} at large scale and low cost, we propose a deep neural-network-based approach built upon commits of open-source repositories.  First, we design and build security patch datasets that include \jingkaicyan{38,291 security-related commits and 1,045 CVE patches from four large-scale C programming language libraries. We manually verify each commit, among the 38,291 security-related commits, to determine if they are security-related.}

We devise and implement a deep learning-based security patch identification system that consists of two composite neural networks: one commit-message neural network that utilizes pretrained word representations learned from \jingkai{our commits dataset}; and one code-revision neural network that takes code
before revision and after revision and learns the distinction on the statement level. Our system leverages the power of the two networks for Security Patch Identification. Evaluation results show that our system significantly \jingkaicyan{outperforms SVM and K-fold stacking algorithms.} The result on the combined dataset achieves as high as 87.93\% F1-score and precision of 86.24\%. 

\jingkaicyan{We deployed our pipeline and learned model in an industrial production environment to evaluate the generalization ability of our approach. The industrial dataset consists of 298,917 commits from 410 new libraries that range from a wide functionalities. Our experiment results and observation on the industrial dataset proved that our approach can identify security patches effectively among open-sourced projects. }
\end{abstract}




\maketitle

\section{Introduction}
\label{sec-intro}
Open-source software (OSS) has remarkably impacted modern software development. The recent years have witnessed impressive increasing rates in both the growth and consumption of open source libraries. For example, in 2018 the number of Java packages doubled to 262,488 in Maven central, a Java open source ecosystem \cite{modulecounts}. With the increasing popularity of OSS, vulnerabilities and exploits have been reported more often. It is reported by the security company Snyk that there is an 88\% increase of reported vulnerabilities over two years \cite{snyk}.

Vulnerabilities can be normally fixed by a security patch. It is time-critical to update security patches as a delay in remediation could expose software systems to attacks. One of the most prominent examples is CVE-2017-5638, a vulnerability in Apache Struts 2~\cite{CVE2}, which allowed attackers to gain control of the webserver and execute arbitrary commands. The notorious exploitation after the public disclosure put 143 million Americans' information at risk in the Equifax mega-breach. \jingkaicyan{To promote software security among developers, GitHub provides services, such as GitHub Security, to inform developers whenever vulnerability dependencies have been discovered in the form of security alerts. Developers can often remediate the security alerts by either upgrading their vulnerable dependency to the non-vulnerable version or removing the vulnerable dependency from their project. As reported by GitHub in 2019, there are about 7.6 million security alerts remediated~{\cite{githubreport2019}}, mostly in the form of commits. } Meanwhile, there are only 12,174 Common Vulnerabilities and Exposures (CVE) reported in the same year \cite{cve_details}. This means that a large number of security issues are not reported in the form of CVEs and silently patched into OSS without public notification. 

Security patches are commonly published to the public for security management purposes. Security companies, e.g., Snyk, and WhiteSource~\cite{whitesource, snyk},  keep tracking of public vulnerability libraries, such as NVD \cite{nvd} and VulnDB \cite{vulndb}, to get the latest vulnerabilities and patches. However,  there is usually a delay between the announcement of security bulletins and the application of the patches to the affected software. Almost 70\% of security vulnerabilities~\cite{largepatch}, in open-source software, are patched before \jingkai{these vulnerabilities are reported in vulnerability databases, such as NVD{~\cite{nvd}} or CVEDetails{~\cite{cve_details}}. Security patches are often fixed without any notification. While NVD is one of the common vulnerabilities databases, the developers can update their security blogs or notify security blogs about their security patches. However, they normally do not do so due to time constraints or simply neglecting the importance of the patches.}
This means that users have no knowledge of the security patches until a later time. Dependency in these unsafe versions of libraries could expose the dependent software to hidden risks. To avoid the exploitation of these unsafe libraries, the security patches shall be identified and pushed to the vulnerable software at the earliest. 

\textbf{Motivation.}\jingkai{ Security patches are released to the public through official channels or vulnerability database. As mentioned previously, almost 70\% of security patches are patched even before the vulnerability is released to the public. Often, security companies, such as SNYK~{\cite{snyk}} and SourceClear~{\cite{sourceclear}}, crawled these public vulnerability databases~\mbox{\cite{nvd, cve_details}} and security bulletin~\mbox{\cite{ossfuzz,bugzilla}} to mine patches, security bugs, and vulnerabilities. However, as open-source software grows in size and complexity, security patches are often silently committed into their codebase without any formal notification.}

\jingkai{Our approach aims to \emph{complement the mining of security patches, both implicit and explicit patches, by employing machine learning into identifying security patches} amid a large amount of commits. Security companies can employ our approach to further enhance their security-related databases, providing much comprehensive coverage for security patches and bugs.} \jingkaicyan{Vulnerability knowledge databases can be served for different purposes, for example, to scan vulnerabilities in software development when developers importing vulnerable APIs from open-source or third-party libraries. In vulnerability management companies, after a vulnerability is identified, further security analysis such as risk score, vulnerable versions, vulnerable code have to been done as well, so that the information can be used by their in-house vulnerability scan tools. For example, by comparing the imported versions of open-source or third-party libraries of a project with its vulnerability knowledge base, the scan tool scans if the project, directly or indirectly, introduces vulnerability risks. However, how to build and scan the dependency of a project is beyond the scope of this work. Besides the commercial application of vulnerability management, the collected vulnerabilities can be used for vulnerability research.} 

\jingkaicyan{ Our approach, and in turn our dataset, provides \emph{security patches on a large scale and could be used in different vulnerability research}, for instance, auto-patching and vulnerabilities detection. Researchers can employ our approach in crawling large amount of security patches regardless of the programming language.} \jingkai{On the other hand, an efficient patch detection mechanism allows developers to be well-informed of any hidden security patches even without any notification, hence, ensuring better security for their systems.}

\jingkaicyan{We provide two use case scenarios that finding security patches that can be useful in practice. (1) Our work can be used to enrich the security patches dataset in the areas of software security research. Some works focus on auto-patching and program repairs, such as~\mbox{\cite{li2020dlfix, deepfix, Dinella2020HOPPITY:}}. These works either use small-scale dataset~{\cite{li2020dlfix}}, crawled git commits simply ~{\cite{Dinella2020HOPPITY:}}, or use simple C programs that are written by students~{\cite{deepfix}}. Higher quality security patches dataset can be crawled using our approaches and hence further improves this area of research. (2) As mentioned previously, security patches are often fixed without any notification and they are not known publicly even to researchers. By compiling a dataset of hidden security patches, developers can find relevant fixes through our dataset of patches to find existing fixes. There are many commercial Software Composition Analysis (SCA) tools that provide patching details for the developers to fix these vulnerabilities in open-source projects, such as SNYK~{\cite{snyk}} and BlackDuck~{\cite{blackduck}}. Our research could enhance these knowledge bases and provide high quality patching information to developers. On the other hand, researchers could also research new ways of repairing programs through our extensive dataset. }

Some recent works are focusing on identifying known vulnerabilities and security patches in open-sourced projects~\cite{8809499, wang2016, ijcai16cnn, ndss18vuldeepecker}. These works mostly directly employ conventional machine learning methods and have not sufficiently explored the semantics of commits and source codes. They utilize either handcrafted features based on domain knowledge or solely on commit messages. These features do not generalize well and may carry biases from domain experts. A relevant work by Wang et al \cite{8809499} involves using handcrafted features, such as the number of memory operators, number of loops, etc., to identify security patches among commits. Some of these handcrafted features, such as the number of memory operators, cannot be applied to other projects. As vulnerabilities vary from projects and languages, it is challenging for human experts to select features manually from \jingkaicyan{large} OSS. 

There are several challenges to apply machine learning techniques in this approach. 
\emph{Our first challenge is to compile a modest dataset labelled with ground truth since there are no public datasets available}. This could require considerable manual efforts on data labelling and dedicated design on the very first step of data gathering. While security patches occupy a tiny portion of the entire commits, immediate collection of all commits leads to the high cost of data labelling, and the issue of extremely imbalanced data. Compared with typical short text classification such as \jingkaicyan{Twitter} and movie reviews \cite{wang2017combining}, commit messages are more lengthy, domain-specific, and noisy, usually including variables, file paths, code snippets (See the example in Table~\ref{tbl-commit-example-1}), and sometimes long lists of log from running tests. Thus, \emph{the second challenge is to build an effective commit-message neural network to learn from noisy and domain-specific commit messages of our compiled dataset with moderate data size.} \emph{The third challenge lies in learning from code revisions}. The emerging approaches \cite{ndss18vuldeepecker, dam2017automatic} of building predictor of vulnerabilities by treating source code as a form of text operate on the coarse source-file level and cannot reflect changes of code fragments. Code revisions consist of fragmentary code from usually multiple source files. It captures the syntax and semantics of source code in a finer granularity, but with limited and complex information to infer sound representations. The implicit distinctions within security-related and unrelated codes make code embedding methods in other tasks such as the general distributed representation of code \cite{c2v-18} inapplicable to obtain valid representations of vulnerabilities.

To address the above challenges, we propose a scalable solution to automatically identify security patches in OSS using deep neural networks. We named our approach as \underline{\textbf{S}}ecurity \underline{\textbf{P}}atch \underline{\textbf{I}}dentifier ({\tool}). 
To overcome the first challenge, we propose a keyword filtering process to avoid extremely unbalanced datasets and facilitate the manual verification process to ensure the data integrity and scalability of the solution.  Furthermore, we collected our dataset from four diversified C Projects that could aid to address the second challenge. \jingkai{These four C Projects spans a wide range of functionalities, \textit{e.g.,} operating system, emulators, multimedia-related libraries suite, and network-related software. Hence, source code from multiple domains is included in our dataset, instead of focusing on a single domain.}
With the diversified dataset, our models can learn from both the domain-specific and generic features from the commit message. We specifically employed a statement-level LSTM model to address the third challenge. As developers often make changes statement by statement, we propose a statement-level LSTM to ensure that our model can effectively learn from code-revisions, as opposed to function-level source code or even file-level source code. In the combination of the two-component networks, our model automatically captures semantics from both commit messages and code-revisions, instead of handcraft features, to determine if the commit is a security patch. 
In summary, our contributions are as follows.

\def\Plus{\texttt{+}}
\def\Minus{\texttt{-}}
\begin{table}
	\begin{center}
		\caption{Commit example from Linux kernel}
		\label{tbl-commit-example-1}
		\begin{tabular}{p{2.1cm}|p{5.9cm}}
			\hline
			Hash & 98051872fd25077d3b106ab3d2b945fa7025c1ef\\
			\hline 
			Date & Thu, 14 Dec 2017 13:03:17 +0100 \\
			\hline
			Author & Lorenzo Bianconi  $<$lorenzo.bianconi@redhat.com$>$ \\
			\hline
			Message &
			Subject: [PATCH] mt76: fix possible NULL pointer dereferencing in
			mt76x2\_mac\_write\_txwi().
			Verify wcid is not NULL before dereferencing the pointer to initialize
			txwi rate\/power info \\
			\hline
			Additive Changes & \Plus\emph{   if (wcid \&\& (rate-$>$idx $<$ 0 $||$ !rate-$>$count)) \{ } \\
			\hline
			Subtractive Changes & \Minus \emph{ if (rate-$>$idx $<$ 0 $||$ !rate-$>$count) \{ } \\
			\hline
		\end{tabular}
	\end{center}

\end{table}

\jingkai{\emph{First}, we propose a filtering-based mechanism to build security patch datasets. We collected 344,519 unlabeled commits from four open-source libraries in C that are requested by our industry collaborator. We then filter out 40,523 commits out of the 344,519 unlabeled commits through our proposed mechanism and manual verification. To ensure data quality, \textbf{\emph{4}} professional security researchers spend \textbf{\emph{600}} man-hours on the manual labelling. Our dataset is available at \mbox{\url{https://sites.google.com/view/du-commits/home}.}}

  
\emph{Second}, we design an automated \textbf{S}ecurity \textbf{P}atch \textbf{I}dentification system based on commits ({\tool}). \jingkai{Each commit consists of a commit-message and its code revision. We employ two deep neural networks to effectively learn the two respective features of the commits to determine whether if it is a security patch. Our approach consists of two major components: a commit-message deep neural network ({\tool}-CM) that utilize and learn from commit message and a code-revision deep neural network ({\tool}-CR) that automatically learn features from changes in commits. To leverage the benefits of both networks, we ensemble them using a weighted combination to form {\tool}. }
  
\emph{Third}, we present an evaluation to validate {\tool}. The integrated system of the two neural networks advances either of the constitutional neural networks. {\tool} achieves as high as an 87.93\% F1-Score as well as a precision of 86.24\%. We gained an average of 9.94\% in F1-Score when compared to the three commit-message baselines over five datasets. Moreover, when compared to code-revision baselines, we gained an average of 59.29\% in F1-Score. \jingkai{We conduct qualitative analysis (Section~{\ref{implicit}}) on the security patches that have no clear indication of fixing the vulnerability in their commit messages to demonstrate the ability of our model to learn from code revision. We refer to these patches as \emph{Implicit Security Patches}. Our analysis shows that 75\% of the implicit security patches can be effectively identified as security patches by our model through their code revisions.}

  
\jingkai{\emph{Fourth}, we deployed {\tool} in the production cycle of our industry collaborator. It classified \emph{298,917} commits from \emph{410} new C libraries, reducing the amount of effort in the manual verification process. 50.54\% (151,080 out of 298,917) of the collected commits are excluded from our production cycle as {\tool} has determined them as non-security patches. Therefore, the performance in the production data validated the usability and effectiveness of our solution. }

\vspace{-0.1in}
\section{Background}
\label{sec-bg}
\subsection{Problem Definition} \label{problem}
We define a commit that fixes existing vulnerabilities in the source code as a \textbf{\emph{security patch}} (SP), and otherwise \textbf{\emph{non-security patch}} (NSP). Table~\ref{tbl-commit-example-1} shows a security patch example from Linux kernel. It indicates a patch for \emph{NULL pointer deference} \cite{npd}. The commit message of the patch contains several projects related and domain-specific words such as \emph{mt76, dereference, wcid, txwi}. A commit records the added and deleted lines of code, which are called \emph{diff} or \emph{code revision}. Table~\ref{tbl-commit-example-1} shows code changes in the affected source file \emph{mt76x2\_mac.c}, including one deleted line and one added line of code. The deleted line begins with the negative sign, while the added line starts with the positive sign. In this paper, we will refer deleted lines as \emph{negative changes} and added lines as \emph{positive changes}.

We focus on \emph{the problem of identifying security patches, i.e, the commits that fix vulnerabilities, through the approach of deep learning.} In this work, we focus on general security patches, instead of a specific type for security patch that fixes a specific type of vulnerability. The identifying process can be considered as a function:
\begin{equation}
    F(x_i, \theta) = p_i 
\end{equation}
that takes input commit $x_i$ and outputs a probability $p_i$ that the commit fixes a vulnerability. The model parameter $\theta$ is what machine learning techniques target to learn, and the commit $x_i$ has to be represented as a $k$-dimensional vector before it is inserted into any machine learning algorithm. Each commit is labeled as \emph{positive} (denoted as 1) if it is a security patch or otherwise, \emph{negative} (denoted as 0). As our problem is a binary classification problem, we use the commonly used cross-entropy loss as the objective function \cite{Goodfellow-et-al-2016},
\begin{equation}
	L(\theta) = -\frac{1}{n}\sum_{i=1}^{n} [y_i\log(p_i)
	+ (1-y_i)\log(1-p_i)]
\end{equation}
where $n$ is the number of training commits; and $y_i$ is the label of commit $x_i$. 
\subsection{Embedding Techniques}
Commit messages and code revisions are written in natural and programming languages. Both are textual information, while classifiers and neural networks work on vectors of real numbers, and best on dense vectors, where all values contribute to defining an object. Hence, we need to encode the raw texts and encodes them into numerical vectors. Embedding is a standard and effective way to transform discrete input objects like words into useful continuous vectors. In natural language processing, a \emph{word embedding} $\mathbf{W}: \text{words} \to \mathbb{R}^n$ is a parameterized function mapping words in some language to $k$-dimensional vectors. Consequently, sentences can be represented by a matrix, where each word is denoted by a vector. e.g.,
{\small{
\begin{eqnarray}
		Fix \to ~  &[& 0.12, 0.05,-0.03, \dots, 0.02 ; \nonumber\\ 
		a \to ~   &~& 0.01, 0.50, -0.10, \dots, 0.11; \nonumber\\
		crash \to ~ &~& 0.06,  0.01, -0.04, \dots, 0.01~] 
	 \nonumber
\end{eqnarray}
}}
Similarly, code snippets can be tokenized into collections of elements (]e.g., symbols) and embedded into numerical vectors. There are various embedding methods, such as one-hot encoding, bag-of-words, term frequency-inverse document frequency\cite{ullman2011mining}, and word2vec (Continuous Bag Of Words (CBOW) and Skip-Gram)  \cite{mikolov2013distributed}. One of the commonly used embedding techniques, word2vec, employs a shallow neural network to model the relations between a word and its context, thus is capable of capturing semantics.

\subsection{Recurrent Neural Networks and LSTM} 
Recurrent Neural Network (RNN) uses internal memory to model long-term dependencies between sequences of data. They are commonly used to learn on sequence of data, such as time-series and natural language processing \cite{lipton2015critical}. LSTM is a special type of gated RNN that uses a memory cell that could remember long-term dependency and a series of gates that control the input and output.
\begin{equation}
	h_{i} = tanh(x_i, h_{i-1})
\end{equation}
For a current output $h_i$, a hyperbolic tangent is performed on the current input, $x_i$ and the previous state $h_{i-1}$. We employ LSTM in our commit message and code-revision learning to learn on the long-term dependencies between sequence of words/tokens. 

\subsection{Convolutional Neural Network} 
Convolutional Neural Network (CNN) has been used on Text Classification problem before and could produce good results \cite{kim2014convolutional, wang2017combining}. CNN has been often used to learn high-level features among words in a sentence through filters and convolutional operation \cite{conneau-etal-2017-deep}.
\begin{equation}
	c_{i} = \sigma(w \dots x_{i:i+k} + b)
\end{equation}
A feature $c_{i}$ is learned by performed convolutional operation over a window of $K$ words, whereas $w$ and $b$ are the weights and bias of the CNN respectively.

\vspace{-0.01in}
\section{Related Work}
\label{sec-rel}

Many early machine learning works use complexity metrics and machine learning algorithms to build predictors for vulnerability~\cite{Shin2008,Shin2011,Zimmermann2010, harer2018automated, chernis2018machine}. These approaches often conduct source code analysis and rely on the domain experts for feature engineering. The attacked problems include but are not limited to predictions of vulnerabilities \cite{ndss18vuldeepecker}\cite{dam2017automatic}\cite{zhou2019devign}, bugs \cite{ijcai16cnn, oopsla18-deepbug}, defections \cite{wang2016, qrs2015defect}, risk incidents \cite{ccs2017riskteller},  code clones \cite{ase16clone}, and various other applications \cite{fse16api, fse18-deeptype, pldi14code, icse18-gu, c2v-18, liu2020atom, liu2020unified, liu2021retrievalaugmented, siow2020core}. 

The emerging approaches treat software code as a form of text and adopt natural language processing techniques to build predictors of vulnerabilities \cite{ndss18vuldeepecker, dam2017automatic, wang2016, li2018vul, pang2017predicting}. These works operate on a coarser level of source files, instead of fragmentary code revisions in a commit. They \cite{ndss18vuldeepecker, dam2017automatic, wang2016, li2018vul, russell2018automated} directly take the source code as the data sources for training neural networks. For example, \cite{drapervul} proposed to use bag of words and CNN to detect vulnerable functions from a dataset labelled by static analysis tools. \jingkai{There are some datasets that are curated by previous works, such as {\cite{fse2014github}} and {\cite{li2018vul}}. We similarly follow their works to compile our datasets{\cite{nvd}}. However, their datasets are much smaller, {\textit{e.g.}} in thousands, hence not suitable for large-scale deep learning. Besides, there are few datasets open-sourced, such as SARD~{\cite{SARD}} used by~{\cite{ndss18vuldeepecker}}. 
These datasets are not vulnerabilities introduced by developers during software development, instead of artificial and highlight the vulnerabilities and fixes in the source code, making it easy for machine learning models to identify them.}

Static analysis tools introduce high false positives, thus the result is not accurate. Li et al. \cite{ndss18vuldeepecker} and Dam et at. \cite{dam2017automatic} proposed to automatically learn representations with LSTM networks. It is done by extracting two specific kinds of code segments with a dependency graph and commercial tool CheckMarx \cite{checkmarx}. These two types of code segments are heuristic indications of vulnerabilities caused by improper uses of library/API functions. Therefore, it will fall short in detecting other types of vulnerabilities. They improved the approach with programming slicing to cover 2 more types of code segments that are related to vulnerability syntax characteristics \cite{li2018vul}. However, both \cite{ndss18vuldeepecker} or \cite{li2018vul} rely on experts to manually extract the most relevant vulnerability syntax characteristic via CheckMarx. 

Two works by Dam et al. \cite{dam2017automatic} and
Wang et al. \cite{wang2016} demonstrated the effectiveness of neural networks in automatic feature learning from source files. A comparison of the two approaches was conducted in \cite{dam2017automatic}, where the result showed that LSTM networks outperform deep belief networks in feature learning. Huo et al. \cite{ijcai16cnn} utilized CNN to determine if a bug report was related to a source file, and Mou et al. \cite{aaai2016cnn} applied CNN on an abstract syntax tree to detect code snippets of certain patterns. \cite{patchnet} designed a tool to classify buggy patches by extracting features from commit messages and code changes. An empirical study has been conducted to investigate the effectiveness of generating fixing patches by learning from buggy commits \cite{Tufano:2018:EIL:3238147.3240732}. The evaluation results showed their model fixed around 9\% of the test cases correctly by generating patches.

\jingkaicyan{Several previous works, such as \mbox{\cite{farsec, secemo, atn}}, mine security keywords based on the vulnerability data and focus on identifying security bug reports and non-security bug reports. 
However, there are essential differences between a security patch and vulnerability, thus the using keywords are different for them. For example, bug reports do not have "fix" or "CVE" in their keywords. Additionally, as we observed, some keywords are library specific, e.g., in Linux, oops are used to describe a vulnerability.
Therefore, we cannot directly apply their keywords to our domain. Our work present an effective and efficient way of processing a large amount of data. Therefore, we choose a keyword search as part of our filtering process. Our keywords have been handpicked by security researchers and proved through CVE coverage in our experiment. Out of a total of 238 Linux security patches, 89.49\% of the patches can be covered by our keywords. This shows that our keywords are sufficient for the approach.}
    
\jingkaicyan{As previous works have shown that the keywords are gathered based on a single dataset, the keyword list also can be expanded iteratively. Therefore, we considered keyword search as one of our future works and focus more on identifying security patches in this paper. More sophisticated methods and techniques can be explored and used in this area of keyword filtering and we will look into that area in our future works.}

\jingkaicyan{There are several other works that focus on finding security artifacts, such as \mbox{\cite{msr2020, seip}}. In one of the work by Yang et al.~\mbox{\cite{msr2020}}, they focus on finding vulnerability-related items that span across JIRA tickets, Bugzilla Reports, etc. While they are locating vulnerability-related items, our work focus on \emph{security patches}, i.e., commits that focus on fixing the vulnerabilities. These are two closely-related but different objectives. For instance, the keyword for vulnerability-related might be different from the keyword for security patches. Another work by Yang et al.~\mbox{\cite{seip}}, has a related but yet different motivation from our approach. It mainly focuses on finding the related library or libraries of vulnerability. Taking CVE-2020-8428 as an example, their work focus on finding the name of the library in CVE-2020-8428, which in this case, is Linux. On the other hand, our work involves finding the patch, i.e., the fixes of CVE-2020-8428. These are two different motivations and approaches. Although both of our objective and motivation is to build a security knowledge base, these are two orthogonal research areas and can be performed together. }

\section{Approach}
\label{sec-app}
\begin{figure}[tp]
	\centering
	\includegraphics[width=1.0\textwidth]{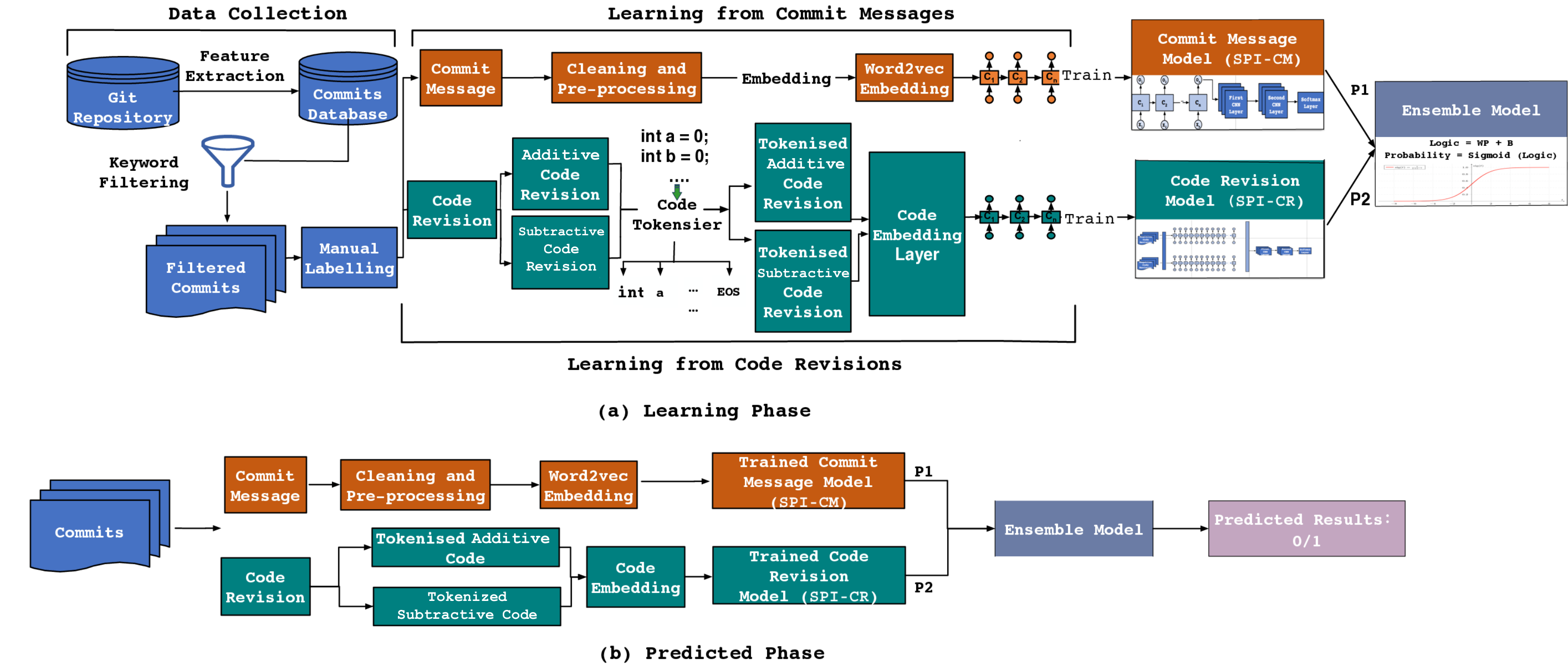}
	\caption{System Overview of \tool}
	\label{fig-network-overal}
\end{figure}

\subsection{Overview}
In this section, we present the design of \tool. It extracts a broad range of security-related data from commits, learns the features from commit messages and code-revisions, and selects high-level features to obtain an informative latent semantic representation. We then efficiently ensemble the unified representations to build a classifier that determines if a commit is a security patch.
Fig.~\ref{fig-network-overal} highlights the framework of \tool system, with part (a) for the training and (b) for the prediction phase. The training phase consists of the following major building blocks.

\begin{itemize}
	\item \emph{Data collection}. Due to the large volume of commits and a notably low rate of security-related commits, we design a keyword filter to exclude the majority of security-unrelated commits. 	
	
	\item \emph{Learning from Commit Messages}. We propose a deep neural network, called \tool-CM, that is built upon the textual information of commit messages to identify security patches among commits.
	
	\item \emph{Learning from Code Revisions}. Similar to \tool-CM, we design two-layers deep neural networks that utilize both subtractive and additive code changes to identify security patches - \tool-CR.
	
	\item \emph{Ensemble Learning}. We ensemble both neural networks, \tool-CM and \tool-CR to form our proposed neural networks that utilize both commit message and code revision, \tool. 	
\end{itemize}

\noindent In the prediction phase, we use the learned models in the training phase to predict if a given commit patches a vulnerability or not.

\subsection{Guiding Principles for Design of Deep Neural Networks}
As there are so many variants of neural networks for different applications and datasets, we need principled guidelines to design effective neural networks to learn from commit messages and code revisions regarding the task of finding security patches.
 
\emph{Commit Message Learning.} Commit messages are mostly in natural languages, though including file paths, code snippets, and various other noise. \emph{\jingkai{The neural networks is required to understand the contexts of a message and to decide whether it describes a security patch.}} Hence it is straightforward to choose LSTM-based networks because they are an effective option to cope with natural language. Besides, the average length of commit messages we collected is 69 by word count, neither too short (e.g., around 10) like Tweets \cite{wang2017combining} that need the extra design to combine more knowledge nor too long (e.g., more than 300) that goes beyond the capacity of LSTM networks. 

\emph{Code Revision Learning.} Learning from code revisions is challenging because the data are in programming languages and the patch characteristics are implicit. Though code is sequential,  it contains structural and logical semantics. Previous work \cite{ndss18vuldeepecker} shows LSTM networks are capable of capturing the context in the source code to some extent. However, \emph{in code revision, it requires the neural networks to tell if it is a security patch from the difference between the source code before and after revisions}. It means we cannot directly feed the whole changed code without differentiating the deleted old code and the added new code to an LSTM network like \cite{ndss18vuldeepecker}, and expect the network to figure out the revision belongs to a security patch. Therefore, we utilize two LSTMs to learn the context from subtractive and additive changes, followed by CNN layers to learn the difference of the two versions statement by statement as CNNs can select higher and useful features related to the targets.
\begin{table}
	\begin{center}
		\caption{Keywords in Filtering Process}
		\label{tbl-keywords}
		\begin{tabular}{p{0.9cm}|p{7.cm}}
			\hline
			Types & Keywords List \\
			\hline 
			General & \textit{out of bound, use after free, double free, divide by zero, overflow, illegal, leak, 
				disclosure, improper, unexpected, sanity check,
				uninitialize, fail, null pointer dereference, 
				null function pointer, crash, corrupt, deadlock, 
				race condition, denial of service, 
				CVE, exploit, attack, vulnerable, fuzz, verify,
				security issue/problem/fix, privilege, malicious,
				undefined behavior, exposure, remote code execution,
				open redirect, OSVDB, ReDoS, NVD, 
				clickjack, man-in-the-middle, hijack, advisory,
				insecure, cross-origin, unauthorized, infinite loop,
				authentication, brute force, bypass, crack, credential, 
				hack, harden, injection, lockout, password, 
				proof of concept, poison, privilege, spoof, compromise, 
				valid, out of array, exhaust, off-by-one, privesc, 
				bugzilla, limit, craft, overrun, overread, 
				override, replay, constant time, mishandle, 
				underflow, violation, recursion, snprintf, 
				initialize, prevent, guard, protect} \\
			\hline
			Library-Specific & \textit{KASAN, general protection fault (GPF), 
				oops, panic, syzkaller, trinity,
				grsecurity, vsecurity, oss-security 
			} \\
			\hline
		\end{tabular}
	\end{center}
	\vspace{-5mm}
\end{table}

\subsection{Data Collection}
\label{sec-data-collection}
We collected our data from two sources, \emph{NVD} and \emph{Manual Verification of Commits.} The patches of CVEs on NVD, if available, can serve as labelled data of security patches. Although these patches are very useful to our data, they are generally very limited. Among our total dataset of 40,523 commits, there are \emph{only 1,045 security patches} that are crawled from NVD. Therefore, we need an alternative source for more security patches. 

\emph{Manual Verification of Commits.} We extracted a wide range of security-related commits from selected libraries by following a two-step process: \emph{Commits Filtering} and \emph{Manual Labelling}. We use four popular and diversified open-source libraries, \emph{i.e.}, Linux, FFmpeg, Qemu, and Wireshark. They are popular OSS from different applications. For example, Linux is an operating system and Wireshark is a popular network packet analyzer. This diversified dataset allows us to generalize our model onto different C libraries. \jingkaicyan{For each project, we gather their commit details, messages, and patches.} Due to a large amount of commits in the Linux repository (750k), only commits from 2016 and 2017 are collected. \jingkaicyan{For FFmpeg, Qemu, and Wireshark, we crawled their commits up to January 2018.}

\noindent\textbf{\emph{Commits Filtering.}} 
\jingkai{Each selected library has at least 55k commits and most of the commits in the selected libraries are not related to any vulnerabilities or security issues. We show the statistics of the collected commits across different projects in Table~{\ref{tbl-dsoverview}}.} There is a need to filter out irrelevant commits to have balanced data. We use regular expression to exclude commits whose messages are not matched with a pre-defined set of security-related keywords \jingkaicyan{(shown in Table~{\ref{tbl-keywords}}). We derived the keywords by manually inspect security patches of the CVEs and extract the common words.} Hence, they are closely related to security or vulnerabilities. \jingkaicyan{Algorithm~{\ref{algo-kw}} shows the pseudo-code for the keyword filtering process. Given a single commit, the algorithm will return True if the commit contains any of the keywords. Otherwise, it will return False}

{
\centering
\begin{minipage}{.7\linewidth}
\begin{algorithm}[H]
  \caption{Keyword Filtering Algorithm}\label{algo-kw}
  \begin{algorithmic}
  \STATE \textbf{Input}: Commit Message $c$, Keyword List $kws$
    \FOR{$word$ in $c$}
      \IF {$word$ in $kws$}
        \RETURN True
      \ENDIF
    \ENDFOR
    \RETURN False
  \end{algorithmic}
\end{algorithm}
\vspace{2pt}
\end{minipage}
\par
}

\jingkaicyan{Despite that we may miss the commits that do not contain security-related keywords, the keyword filtering process is still an effective process to filter out security patches. We validate our keyword filtering process with 238 Linux commit patches from 2016 and 2017. These commit patches are fixes for official Linux CVEs that we are able to find and crawl from vulnerability databases, such as NVD. These patches are standard vulnerability patches including commit messages with and without keywords. According to our validation, our keywords can cover 89.49\% of the CVE commits. Thus, the majority of the patches are captured by our keyword filtering process. Furthermore, our experimental results in Section 6.2 shows that our model can correctly predict implicit security patches with high accuracy. }

Table~\ref{tbl-keywords} includes the main keyword list (not including derived forms). We separate them into two main categories: library-specified and general keywords. Some libraries are more prone to specified types of vulnerabilities or use some specific words/tools to name/discover them, such as \textquotesingle oops\textquotesingle  ~or \textquotesingle KASAN\textquotesingle  ~in Linux. We observed that keywords that are related to memory vulnerability are matched more often in our dataset. Keywords, such as \emph{null pointer dereference}, \emph{uninitialize/initialize}, \emph{overflow}, \emph{corrupt}, appeared in the top matched keywords. These keywords often appear in memory-related vulnerability, such as null pointer exception \cite{npd}, uninitialized variable \cite{uv}, and buffer overflow \cite{bo}. We present the top matched keywords for several keywords in Table \ref{tbl-keywords-count}. We omitted the frequency count for generic words, such as \emph{crash}, \emph{check}, and \emph{fail}. 

\begin{table}[t]
	\centering
	\addtolength{\tabcolsep}{-3pt}
	\caption{Frequency of Keywords}
	\label{tbl-keywords-count}
	\begin{tabular}{|l| c c |}
		\hline
		\textbf{Keywords} & \textbf{Count} & \textbf{Ratio}  \\
		\hline
		uninitialize 	& 4,532 & 0.111 \\
		\hline
		infinite loop & 3221 & 0.0794 \\
		\hline
		overflow 	& 2,577 & 0.0635  \\
		\hline
		race-condition & 2274 & 0.0561 \\
		\hline
		out-of-bound & 1378 & 0.0340 \\
		\hline
		null pointer dereference & 1,294 & 0.0319  \\
		\hline
		corrupt	& 1,228 & 0.0303  \\
		\hline
		deadlock & 805 & 0.0198  \\
		\hline
        use-after-free & 735 & 0.0181 \\
        \hline
		sanity check & 628 & 0.0154  \\
		\hline
	\end{tabular}
	\vspace{-4mm}
\end{table}

Finally, we remove those commits with empty code-revisions and/or commit messages. Table~\ref{tbl-dsoverview} summarizes the statistics of the keywords filtering process. Notwithstanding Linux has \jingkaicyan{large} amount of commits, just 7.4\%  are left after the filtering phase. The keyword filtering phase also filters the other three libraries to 11\% to 18\% of its total commits.

\noindent\textbf{\emph{Manual labelling.}} 
We employed four experienced security researchers to manually label the filtered commits. Our four security experts consist of 3 bachelors and 1 Ph.D., all of whom are hired by the industrial collaborator and have sufficient experience of reporting their CVEs to NVD. \jingkai{Each of the bachelor evaluators has at least 2 years of security-related experience and the Ph.D. evaluator have at least 3 years of security-related experience. One of our authors participates in labeling of the filtered commits.}
Since some commits might not have a clear indication of fixing a vulnerability in their messages, it is difficult to distinguish a security patch solely from their commit messages. Hence, the process requires security researchers to identify security patches through their code revisions. To reduce the number of false positives and ensure the integrity of our data, we exercise a two steps manual labelling process:
1) Initial verification. Each filtered commit is examined and labelled by two different security researchers \jingkai{independently} into one of the three categories: security patch, non-security patch, unsure.
2) Final confirmation. If the two initial labels of a commit are different or include unsure labels, the commit would be forwarded to a senior researcher for further investigation. Commits that are labelled with unsure will be excluded from the dataset.

\noindent\textbf{\emph{Ground Truth.}}
The ground truth consists of the security patches of CVEs and manually labelled security patches, \emph{i.e} commits that fix vulnerability, gathered from the previous steps. The total amount of filtered commits with labels is \textbf{40,523}, and it costs around \textbf{600 man-hours} for the manual labelling. 

Table~\ref{tbl-dsoverview} shows the number of security patches and non-security patches for each project. We discovered that Linux has about 69.4\% commits that are security patches, a much higher rate compared to the other three libraries. The other three libraries, i.e., FFmpeg, Qemu, and Wireshark correspondingly have 45.7\%, 43.0\%, and 41.5\% commits labelled as security patches. As we performed experiments within each project, each project dataset is randomly spilt
into training (75\%) and testing set (25\%) for their respective evaluation. Furthermore, we conduct experiments on the combined dataset, i.e., the dataset that contains commits of all four projects. This complete dataset will also be split into training (75\%) and testing set (25\%) for training and evaluation of the baseline and models.

\subsection{Learning from Commit Messages}
\label{subsec-msg}

\begin{figure*}[t]
	\begin{minipage}{0.45\textwidth}	
	\centering
	\includegraphics[width=0.95\linewidth]{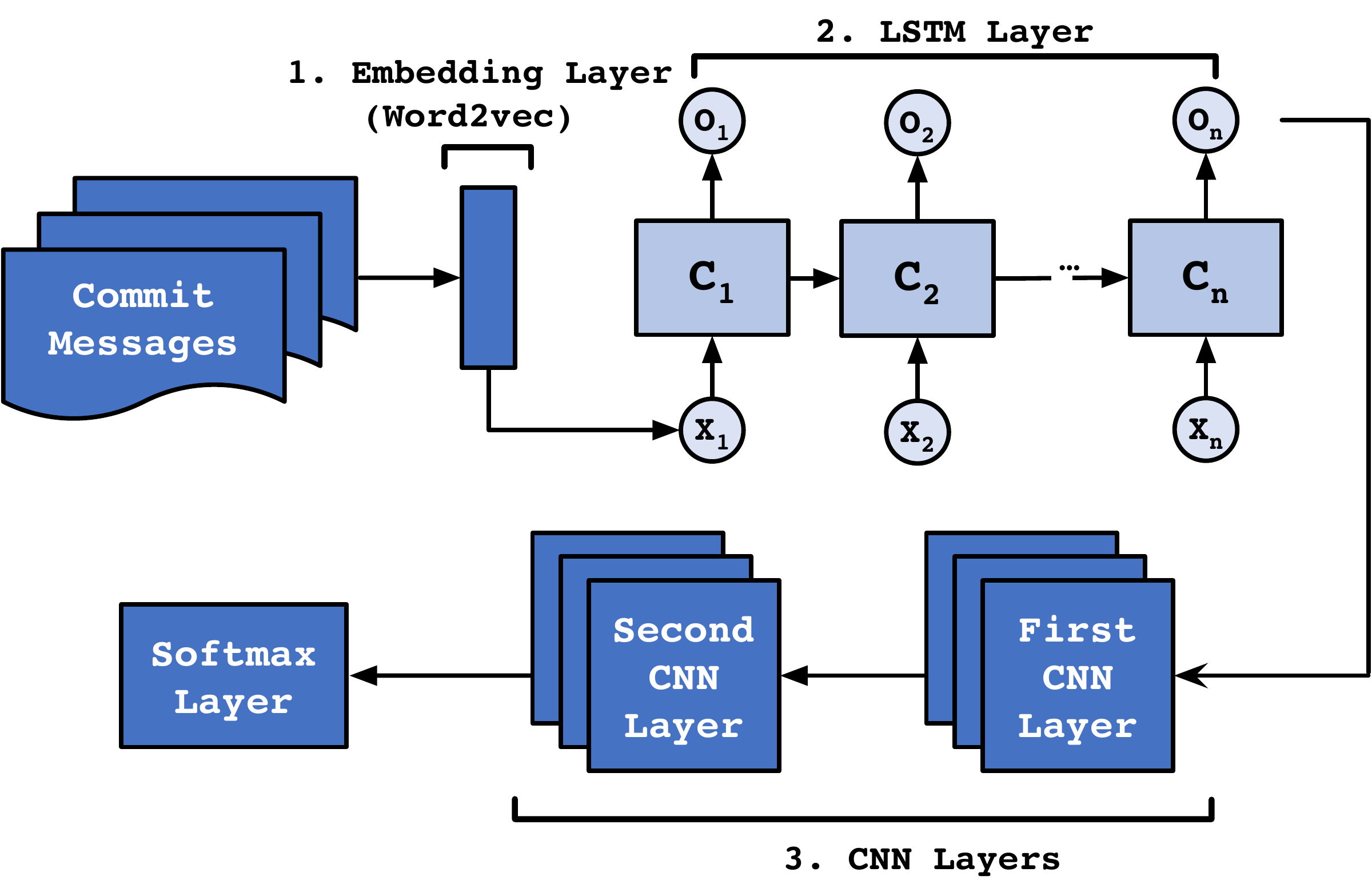}
	\caption{Deep neural network for commit messages }
	\label{fig-network-msg}
	\end{minipage}
	\hfill
	\begin{minipage}{0.45\textwidth}
	\centering
	\includegraphics[width=1\linewidth]{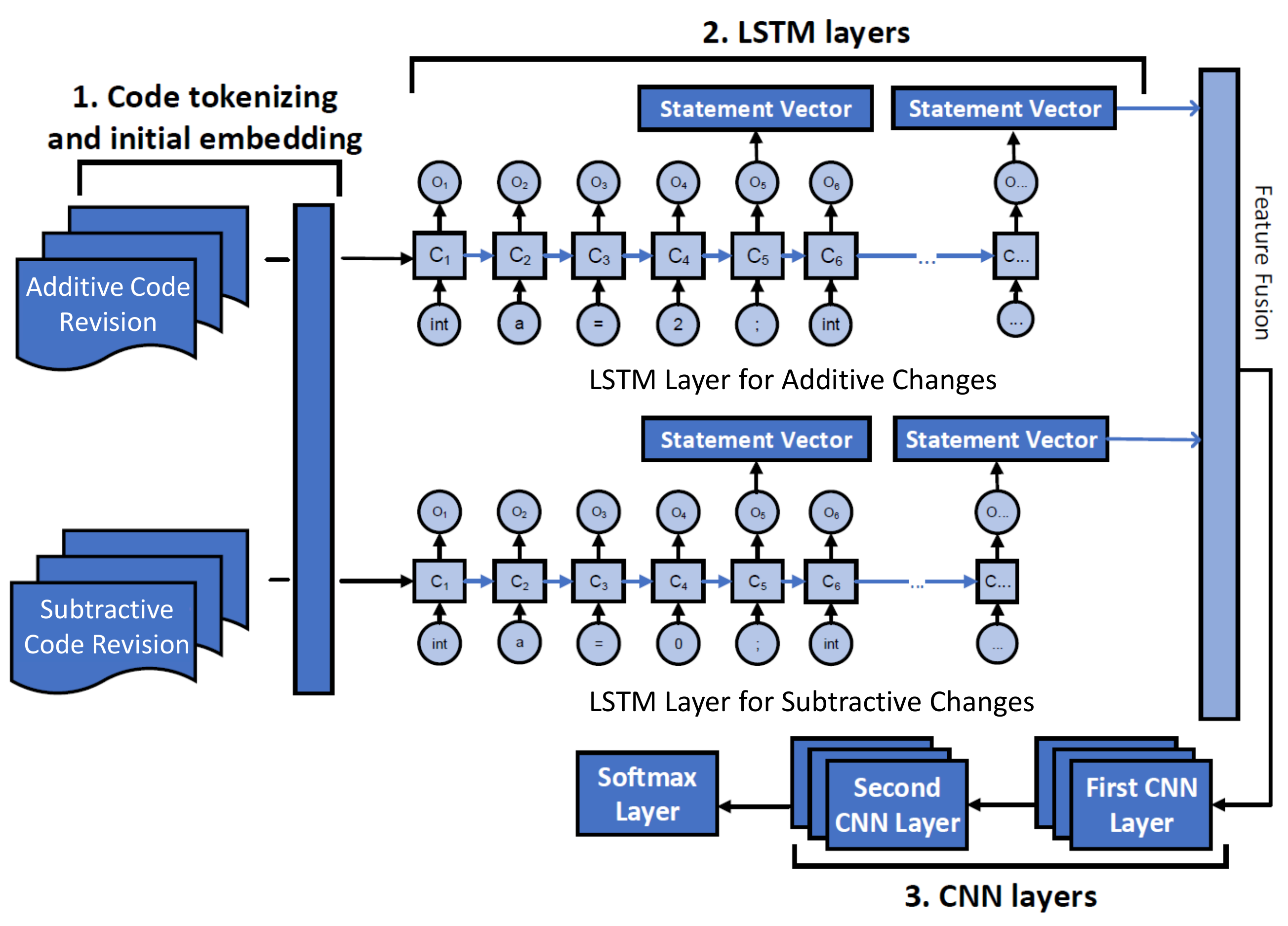}
	\caption{Deep neural network for code revisions}
	\label{fig-network-diff}
\end{minipage}
\hfill
\end{figure*}

Fig.~\ref{fig-network-msg} shows the proposed network architecture for learning from commit messages. 
This component comprises these four parts sequentially.

\emph{1. An embedding layer to represent tokenized words of input messages with numerical vectors}. To feed the raw texts to LSTM layers for feature learning, we firstly encode commit messages in the input layer. We pretrain a word2vec model with all commit messages that are not filtered by the keywords. Utilizing pretrained word2vec models over massive datasets is proved to be more effective in text classification \cite{fse2017vul}. The trained word2vec model consists of $V$ embedding vectors,  where $V$ is the vocabulary size and each vector with $k$ dimensions is a numerical representation of a word. Let $L$ be the maximum length of commit messages. We encode each commit message, using word2vec embedding, into a matrix with a dimension $(L,k)$. The vocabulary size of the pre-trained word2vec model is $283,146$.
	
\emph{2. An LSTM layer to extract representations of commit messages.} The embedded commit message is passed to the LSTM layer to attain middle-level semantic features, where LSTM is designed to capture rich context (\emph{i.e.}, long-term dependency) of words. The output of LSTM is inputted into convolutional layers for higher-level features. To ease the tuning of hyperparameters, we opt to use one layer of LSTM. Let $N$ be the number of units in the LSTM layer, the output of the layer for a single commit message becomes a $(L, N)$ vector.
	
\emph{3. CNN layers to select higher-level representations.} Convolutional layers aim at electing abstract representations from the output of LSTM. Contrasted to the predictors that immediately pass the learned features of LSTM to a softmax layer for classification, we noticed that the predictors with CNN layers perform better. In our evaluation, we practice 2 convolutional layers, each followed by ReLu activation functions. The output of the CNN layers is, consequently, a one-dimensional vector.

\emph{4. Softmax layer to output the predicted probabilities.} Learned features by the previous CNN layers are directly passed to a 2-unit softmax output layer to determine the likeliness that a commit message indicates the commit fixed a vulnerability or not, \emph{i.e.}, a security patch.

\begin{figure}[t]
	\centering
	\includegraphics[width=0.45\textwidth]{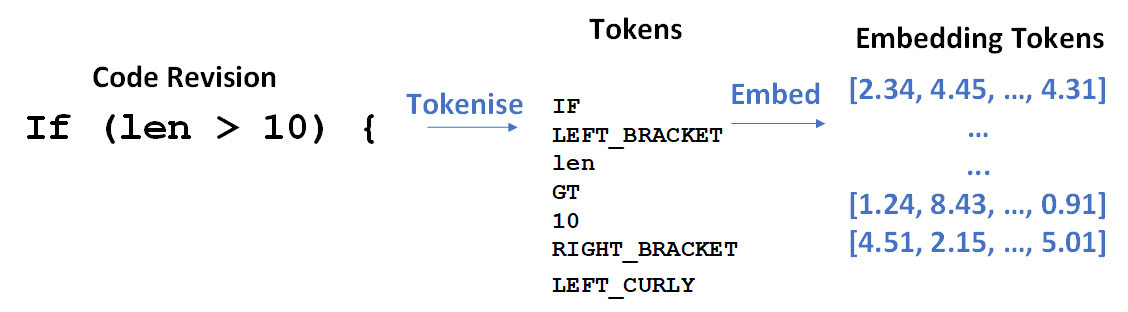}
	\caption{Example of code tokenizing and embedding  }
	\label{fig-token-process}
	\vspace{-5mm}
\end{figure}

\begin{table}[t]
	\centering
	\addtolength{\tabcolsep}{-3pt}
	\caption{Overview of datasets }
	\label{tbl-dsoverview}
	\begin{tabular}{|l| c c c c c |}
		\hline
		\textbf{Project} & \textbf{Total} & \textbf{Filtered} & \textbf{CVE} & \textbf{Filtered Ratio} & \textbf{SP Ratio} \\
		\hline
		\textbf{Linux} 	& 144601 & 10731 & 238 & 0.0742 & 0.6948 \\
		\hline
		\textbf{FFmpeg} 	& 80882 & 12123 & 249 & 0.1498 & 0.4578 \\
		\hline
		\textbf{Qemu} 	& 55070 & 10128 & 242 & 0.1839 & 0.4306 \\
		\hline
		\textbf{Wireshark} 	& 63966 & 7541 & 316 & 0.1178 & 0.4155 \\
		\hline
		\textbf{Total} & 344519 & 40523 & 1045 & 0.1176 & 0.5059 \\
		\hline
	\end{tabular}
	\vspace{-4mm}
\end{table}

\subsection{Learning from Code Revisions}
\label{subsec-diff}

To circumvent introducing lengthy sequences of code and reflect the essential changes, we only consider the removed source code before revision (\emph{i.e.}, the subtractive changes) and appended source code after revision (\emph{i.e.}, the additive changes). We employ a more dedicated neural network for learning on code revisions. Fig.~\ref{fig-network-diff} depicts the network architecture.  \jingkaicyan{Both code revision of security-related and security-unrelated commits are used as input to the deep learning model to distinguish security patches and non-security patches.}

\emph{1. Code tokenizing and initial embedding}. 
The code2vec embedding layer similar to the word2vec embedding layer of commit message neural network, where each token of code statements is interpreted by a high-dimensional vector. The code embedding vector is learned by training a word2vec~\cite{mikolov2013distributed} neural networks over the dataset. We split additive and subtractive code revisions into individual statements of code tokens before inputting them into neural networks. We use Pygments~\cite{pygments} to parse the code into the token-level, followed by a text cleaning process. These tokens are a set of variable names, function names, integer literals, logical operators, and C language keywords. To avoid an excessive vocabulary set, we replaced strings like memory addresses and IP addresses with constant strings (e.g., $<INT>, <STRING>$). The entire vocabulary size of all the code revisions is $V = 111,987$. Fig.~\ref{fig-token-process} instances the tokenizing and code embedding process of one code revision. 

\emph{2. Statement-level LSTM layers}.
Code revisions are scattered statements across multiple source files, instead of complete function blocks in one file. Therefore, we design a specific network to learn the semantic meaning of code revisions on a statement-level. To achieve this, in the LSTM layer, our model incorporates two LSTM networks to encode the sequences of code revisions, respectively for additive changes and subtractive changes. We output representations of each statement in the LSTM layers to learn the structural information of source code. Recall that every statement of the input sequences is a sequence of tokens, marked by End of Statement (EOS) at the end of each statement. To obtain a single vector representation per statement, we sample the representation sequences yielded by the encoders exclusively at time steps corresponding to EOSs. 
During the training, the network learns to generate meaningful representations at the position of the EOS inputs. Let $M$ and $P$ be the number of statements in additive and subtractive revision. The results of the LSTM encoders are $M$ representation vectors output by the additive revision encoding LSTM, denoted by $\{v_j , j \in (1,...,M )\}$, and $P$ representation vectors output by the subtractive revision encoding LSTM, denoted by $\{u_i, i\in(1,...,P )\}$. The statement representation vectors are then fused into a matrix.

\emph{3. CNN layers}. We use a multi-layer CNN to extract the difference between the code before and after the revision statement by statement. The output of the CNN layers is inputted into a 2-unit softmax layer for security patch identification.

\subsection{Ensemble Learning}

Through the above feature learning process, we obtain predicted probabilities from both commit messages and code revisions. Our final step  KKH is to consolidate the learned results from commit messages and code revisions. There are multiple ways to achieve this goal. One alternative is to construct a heterogeneous neural network that unites the commit-message and code-revision neural network into one unified network. Nevertheless, we observed it became notoriously tricky to tune the hyperparameters with good performance. Because the parameters we tuned beforehand in the two separate networks are invalid in the fresh setting, and the network complexities increase dramatically. Consequently, we turn to a simpler yet efficient method, ensemble. We obtain a weighted combination of both commit-message neural network output and code-revision neural network output as the ensemble model to output the final classification decision. \jingkai{After tuning the weight in our experiments, we opt to use 0.5 as our value of the weight as it allows us to achieve the best results. However, the value of the weight can be learned and adjusted according to different settings, such as datasets, and employing neural networks. }

\subsection{Security Patch Prediction}

After the learning phase, we obtained three models accordingly for prediction. 

\emph{Security Patch Identifier on Commit Messages (\tool-CM).} We use a 2-unit softmax layer to classify the learned features from Section~\ref{subsec-msg}. It helps us to justify if the deep learning approach performs better than the traditional machine learning algorithms at practically moderate-size datasets. The previous work \cite{fse2017vul} encoded the commit messages with a pretrained word2vec model, and trained an ensemble learning algorithm that leverages several different basic classifiers to achieve the state-of-the-art result. The comparison with this work helps answer the question. 

\emph{Security Patch Identifier on Code Revisions (\tool-CR).} Similarly, the learned features from Section~\ref{subsec-diff} are passed to a 2-unit softmax output layer. Through this, we investigate if we can extract characteristics of meaningful patches from code revisions. 

\emph{Security Patch Identifier on commit messages and code-revisions (\tool)}. We ensemble the results of both \tool-CR and \tool-CM to complement their disadvantages. We then evaluate their performance to see if it perform better.

Fig.~\ref{fig-network-overal}(b) illustrates the process of how a commit is predicted using our deep learning model. Given a commit, its message and code revision is processed respectively and then predicted via  \tool-CR and  \tool-CM. The upper part of Fig.~\ref{fig-network-overal} (b) shows the procedure of \tool-CM: 1)firstly, the message is cleaned and tokenized into lists of words, 2) each tokenized word is embedded by the pretrained word2vec model, so that the commit message is represented by a matrix and ready for input into the trained \tool-CM model, 3) the \tool-CM predictor calculates based on the input embedded matrix and outputs a probability (P1 in the Fig.~\ref{fig-network-overal}(b)).
Similarly, code revision is firstly preprocessed, \emph{i.e.}, split into additive changes and subtractive, tokenized,  and then embedded using the code embedding approach, before inputting into the trained \tool-CR model. Next, the \tool-CR model predicts with an output probability (P2 in the Fig.~\ref{fig-network-overal}(b)).
Lastly, the ensemble model computes a final decision based on P1 and P2 using the learned weight in the training phase.

\subsection{Evaluation Metrics}
\jingkai{We employ three evaluation metrics, \textit{precision}, \textit{recall}, and \textit{F1-Score} as our evaluating metric for our experiments. As mentioned in Section~{\ref{problem}}, we formulate our problem as binary classification. Therefore, these three metrics are appropriate for evaluating our approach. }

\jingkai{Precision measures the proportion of the true positives among the predicted positives. It allows us to gauge on the amount of false positives that our model is predicted. Therefore, a high precision will infer that our model performs well in identifying security patches. The formula of precision is shown in Equation~{\ref{eqn:precision}}}
\begin{equation}
\label{eqn:precision}
    \begin{split}
         Precision = \frac{True Positives}{True Positives + False Positives}
    \end{split}
\end{equation}

\jingkai{In contrast, recall measures the total number of true positive over actual positive. If the recall of our model is high, it implies that our model is also effectively in determining non-security patches. Recall can be computed as shown in Equation~{\ref{eqn:recall}:}}
\begin{equation}
\label{eqn:recall}
    \begin{split}
         Precision = \frac{True Positives}{True Positives + False Negatives}
    \end{split}
\end{equation}

\jingkai{F1-Score is commonly used in binary classification and previous works~\mbox{\cite{li2018vul, fse2017vul}}. It is computed using a weighted combination of precision and recall. A high F1-score implies the model has a low number of false positive and false negatives. F1-score can be computed using an equation shown below:}
\begin{equation}
    \begin{split}
         F1 = 2*\frac{precision*recall}{precision+recall}
    \end{split}
\end{equation}

\section{Evaluation}
\label{sec-eval}
To assess \tool, we conduct experiments on the four project datasets and one production dataset and compare with the start-of-the-art methods. Our experiments aim to answer the following research questions (RQs).
\begin{enumerate}
	\item  \emph{What is the performance, in terms of F1-score, recall, and precision, of \tool-CM? }
	Can \tool-CM learn from domain-specific commit messages, and outperform the state-of-the-art method \cite{fse2017vul}? (Section~\ref{ss-rq1})
	
	\item \emph{What is the performance, in terms of F1-score, recall, and precision, of \tool-CR?}
	Can \tool-CR learn useful representations from code revisions for identification of security patch? (Section~\ref{ss-rq2})
	
	\item \emph{What is the performance, in terms of F1-score, recall, and precision, of \tool?}
	We evaluate {\tool} and observe if it can perform better than any of the single component. (Section~\ref{ss-rq3})
	
	\item \jingkai{\emph{What is the performance, in terms of F1-score, recall, and precision, of \tool-CM, \tool-CR, and {\tool} in cross-project evaluation?}
	We evaluate our approaches and test the generalization ability of our approach through a cross-project evaluation. (Section~{\ref{ss-rq4}})}
	
	\item \emph{Does {\tool} performs well in production setting?}
    We evaluate if {\tool} works well in industrial production setting. (Section~\ref{ss-rq5})
    
    \item \jingkai{\emph{Does word embedding dimensions and LSTM dimensions affect the performance of {\tool}?}
    We investigate if these hyper-parameters affects the performance of the {\tool} and explore the best value of these parameters. (Section~{\ref{ss-rq6}})}
    
\end{enumerate}

\begin{table*}[t]
\centering
\addtolength{\tabcolsep}{-4pt}
    	    \caption{Evaluation results on Linux, FFmpeg, Qemu, and Wireshark datasets}
    	    \label{tbl-result}
    	    \begin{tabular}{|l| c| c| c | c | c |c | c | c | c | c | c | c |}
		    \hline
			     Method &  \multicolumn{3}{c|}{\textit{Linux}} & \multicolumn{3}{c|}{\textit{FFmpeg}} & \multicolumn{3}{c|}{\textit{Qemu}} & \multicolumn{3}{c|}{\textit{Wireshark}} \\
			     
			     \cline{2-4}\cline{5-7}\cline{8-10}\cline{11-13}
			     & Precision & Recall & F1
			     & Precision & Recall & F1
			     & Precision & Recall & F1
			     & Precision & Recall & F1 \\
			\hline
			   {$K$fs (msg)-RQ1} & 0.8135 & 0.9200 & \textbf{0.8635}
			           & 0.8729 & 0.8119 & 0.8413
			           & 0.7734 & 0.6422 & 0.7017
			           & 0.7932 & 0.6900 & 0.7380
			           \\
		        \emph{LSTM (msg)-RQ1} & 0.8107 & 0.9171 & 0.8606
		                    & 0.8867 & 0.9037 & 0.8951
		                    & 0.8289 & 0.8041 & 0.8163
		                    & 0.8091 & 0.8337 & 0.8212
		                    \\
		        \emph{\tool-CM-RQ1} & 0.8174	& 0.9066 & 0.8597
		                                   & 0.9266 & 0.9187 & \textbf{0.9226}
		                                   & 0.8137 & 0.8277 & \textbf{0.8206}
		                                   & 0.8174	& 0.8750 & \textbf{0.8452}
		                                   \\
		   \hline
		        {$K$fs (code)-RQ2} & 0.7396 & 0.9479 & \textbf{0.8309}
		                & 0.6597 & 0.4805 & 0.5560
		                & 0.6305 & 0.4009 & 0.4901
		                & 0.5950 & 0.3392 & 0.4321
		                \\
		       \emph{LSTM (code)-RQ2}  & 0.7966 & 0.7088 &	0.7502
		                    & 0.5560 & 0.5267 & 0.5409
		                    & 0.4896 & 0.5766 & 0.5295
		                    & 0.4969 & 0.4951 &	\textbf{0.4960}
		                    \\
		        \emph{LSTM (LineReps)-RQ2}  & 0.7598 & 0.7457 & 0.7527
		                               & 0.5135	& 0.6343 & 0.5676
		                               & 0.4319	& 0.4977 & 0.4625
		                               & 0.4788	& 0.4538 & 0.4660
		                               \\
		        \emph{\tool-CR-RQ2} 	& 0.7720 & 0.7125 & 0.7410
		                                    & 0.5126 & 0.6764 & \textbf{0.5832}
		                                    & 0.4992 & 0.6137 & \textbf{0.5506}
		                                    & 0.4807 & 0.4842 & 0.4824
		                                    \\
		    \hline
		        \emph{\tool-RQ3} 	& 0.8788 & 0.9298 & \textbf{0.9036}
		                                    & 0.9652 & 0.9501 & \textbf{0.9576}
		                                    & 0.9440 & 0.9211 & \textbf{0.9324}
		                                    & 0.9016 & 0.8975 & \textbf{0.8995}
		                                    \\
		    \hline
		\end{tabular}
\end{table*}

\begin{table*}[t]
\centering
\addtolength{\tabcolsep}{-3pt}
    	    \caption{Evaluation results on Combined datasets}
    	    \label{tbl-combined-result}
    	    \begin{tabular}{|l| c| c| c|}
		    \hline
			     Method &  \multicolumn{3}{c|}{\textit{Combined}} 
			     \\  \cline{2-4}
			     & Precision & Recall & F1 \\
			\hline
			   {$K$fs (msg)} &  0.7988 & 0.8002 & 0.7995 \\
		        \emph{LSTM (msg)} & 0.8403 & 0.9025 & 0.8703  \\
		        \emph{\tool-CM} & 0.8685	& 0.8866 & \textbf{0.8774}  \\
		   \hline
		        {$K$fs (code)} & 0.6681 & 0.6586	 & \textbf{0.6633} \\
		       \emph{LSTM (code)}  & 0.5959 & 0.7504 & 0.6643 \\
		        \emph{LSTM (LineReps)}  & 0.5488	& 0.8513 &  0.6674\\
		        \emph{\tool-CR} 	& 0.5581 & 0.8656 & \textbf{0.6787} \\
		    \hline
		        \emph{\tool} 	& 0.8624 & 0.8968 & \textbf{0.8793} \\
		    \hline
		\end{tabular}
\end{table*}

\begin{table*}[t]
\centering
\addtolength{\tabcolsep}{-3pt}
    	    \caption{Evaluation results on Cross Projects Evaluation}
    	    \label{tbl-cross-result}
    	    \begin{tabular}{|l| c| c| c| c| c| c| c| c| c| c| c| c|}
		    \hline
			     Method &  \multicolumn{3}{c|}{\textit{LinuxTest}} & \multicolumn{3}{c|}{\textit{FFMpegTest}} & \multicolumn{3}{c|}{\textit{QemuTest}}  & \multicolumn{3}{c|}{\textit{WiresharkTest}}
			     \\
			     
			     \cline{2-4}\cline{5-7}\cline{8-10}\cline{11-13}
			     & Precision & Recall & F1 & Precision & Recall & F1 & Precision & Recall & F1 & Precision & Recall & F1\\
			\hline
			   {$K$fs (msg)} &  0.7657 & 0.7953 & 0.7802 
			                 &  0.7319 & 0.8146 & 0.771
			                 &  0.6696 & 0.6797 & 0.6746
			                 &  0.7395 & 0.6579 & 0.6963 \\
		        \emph{LSTM (msg)} & 0.7048 & 0.9867 & \textbf{0.8222} 
		                          & 0.7834 & 0.8910 & 0.8337
		                          & 0.6886 & 0.7255 & 0.7066
		                          & 0.7987 & 0.8292 & 0.8137 \\
		        \emph{\tool-CM} & 0.7109 & 0.9719 & 0.8211
		                        & 0.8329 & 0.9146 & \textbf{0.8718}
		                        & 0.7008 & 0.7606 & \textbf{0.7295}
		                        & 0.8225 & 0.8063 & \textbf{0.8143} \\
		   \hline
		        {$K$fs (code)} & 0.7918	& 0.2943 & 0.4292
		                       & 0.4988	& 0.6957 & 0.5810
		                       & 0.4702	& 0.7248 & 0.5704
		                       & 0.5064 & 0.4256 & 0.4625 \\
		       \emph{LSTM (code)}  & 0.7301 & 0.8672 & 0.7927
		                           & 0.4844	& 0.9142 & 0.6333
		                           & 0.4394	& 0.9630 & 0.6035
		                           & 0.4208	& 0.9684 & 0.5867 \\
		        \emph{LSTM (LineReps)}  & 0.7213 & 0.8581 & 0.7838 
		                                & 0.4753 & 0.9383 & 0.6310 
		                                & 0.4507 & 0.9277 & \textbf{0.6067}
		                                & 0.4272 & 0.9160 & 0.5827 \\
		        \emph{\tool-CR} & 0.7017 & 0.9902 & \textbf{0.8213}
		                        & 0.4721 & 0.9636 & \textbf{0.6337}
		                        & 0.4434 & 0.9585 & 0.6063
		                        & 0.4214 & 0.9763 & \textbf{0.588}7 \\
		    \hline
		        \emph{\tool}    & 0.7108 & 0.9814 & \textbf{0.8245}
		                        & 0.8304 & 0.9126 & \textbf{0.8733}
		                        & 0.6930 & 0.7714 &	\textbf{0.7301}
		                        & 0.7933 & 0.8477 & \textbf{0.8196} \\
		    \hline
		\end{tabular}
\end{table*}

\subsection{Implementation Details}
We implemented the model in Tensorflow \cite{tensorflow2015-whitepaper}, and Python 3.6. We conducted the training and testing on a server with 36 2.30GHz Intel Xeon processors with a single RTX 2080. We ran our model with a patience of 200 epoch, with Adaptive Moment Estimation (Adam) optimizer \cite{kingma2014adam}, batch size=$64$, learning rate=$1e^{-4}$ and dropout=0.2.

We standardize all our word/code embedding to a dimension of 300.
1) LSTM layer. We use 1 layer of LSTM and 64 LSTM units in all cases. \jingkaiorange{We refer readers to RQ6 as we perform hyper parameter tuning of word dimension and LSTM in RQ6.} We limit the maximum length of the input sequence to be 100. \jingkai{If the length of the input sequence is more than 100, we trim the sequences to be the length of 100. Similarly, if the length of the input sequence is less than 100, we pad the sequences with zero at the end of the vector until the length reaches 100.} We have set the number of statements, for the code-revision models, to be the average number of statements, $10$.  2) Convolutional network layers. We use a 2-layer convolutional network and max-pooling for all experiments. For \tool-CM, we have the first CNN with $64$ filter,  $(3\times 64)$ kernel size, $(1\times 1)$ stride, followed by a max-pooling with $(2\times2)$ pool size and $(2)$ stride, and the second CNN with $32)$ filter, $(3\times 64)$ kernel size, $(1\times1)$  stride, and a max-pooling with $(2\times2)$ pool size and $(2)$ stride. For \tool-CR, both CNNs have $(32)$ filter, $(3\times 64)$ kernel size, and $(2)$ stride.

\subsection{Evaluation Metrics}
To measure prediction results, we ultilize the three metrics commonly used in vulnerability identification and defection identification \cite{ndss18vuldeepecker, dam2017automatic,fse2017vul, wang2016}: \emph{Precision}, \emph{Recall},  and \emph{F1 score}.

\subsection{Baselines}
The most related works are \cite{perl2015vccfinder} and \cite{fse2017vul} that respectively find vulnerability-introducing and vulnerability-fix commits, i.e, security patches. The approach of \cite{perl2015vccfinder} utilized SVM while \cite{fse2017vul} designed a $K$-fold stacking algorithm that ensembles $6$ different conventional classifiers including SVM for better performance and implemented in production. 
Consequently, we directly compare with the $K$-fold stacking algorithm (abbrev. \emph{$K$fs}). We compare \tool-CM and \tool-CR with their respective \emph{$K$fs} baseline and base LSTM model. For \tool-CR, we conduct an additional evaluation to evaluate if the statement-level model, LSTM (LineReps), performs better than basic LSTM. Code-revision commonly consists of statement-level changes, hence, we test our performance of the model on statement-level without CNN to test the performance over basic LSTM. We implemented the baselines and tuned them at the best parameters. We carry the experiments over each project, as well as the combined dataset of the four projects. We split each dataset into training (75\%) and testing (25\%) set for training and evaluation of the models.  The results for each dataset and method are summarized in Table~\ref{tbl-result} and Table~\ref{tbl-combined-result}.   

\subsection{Performance of \tool-CM (RQ1)}
\label{ss-rq1}

We compare our commit-message neural network, \tool-CM, with the $K$-fold stacking algorithm, \emph{i.e}, $K$fs (msg), and LSTM network, \emph{i.e}, LSTM (msg). Table~\ref{tbl-result} presents the three metric scores for learning on commit messages. Overall, it shows that \tool-CM outperforms the $K$-fold stacking algorithm and the basic LSTM network. 

Particularly, compared with the $K$-fold stacking, the increment of the F1-score in \tool-CM ranges from \textbf{7.7\%} to \textbf{11\%} in all experiments, excluding Linux. We noticed that the F1-score for $K$fs (msg) in Linux perform slightly better than \tool-CM, by \textbf{0.38\%}. One possible explanation is that Linux has the most imbalanced data and hence, could affect the results on $K$fs (msg).
Compared with the LSTM network, the F1-score improves \textbf{2.7\%} in FFmpeg, \textbf{0.43\%} in Qemu, \textbf{2.4\%} in wireshark and \textbf{0.7\%} in the combined dataset.  

The evaluation proves the effectiveness of CNNs in our design to obtain better features. We also observe that the performance on the combined dataset is generally better than that of a single project, indicating that big data volume boosts the performance.

\begin{tcolorbox}[breakable,width=\linewidth,boxrule=0pt,top=2pt, bottom=2pt, left=1pt,right=1pt, colback=gray!20,colframe=gray!20, ]
\textbf{Answer to RQ1:} Our proposed deep neural network {\tool-CM} proved advantages over \jingkaicyan{the comparing baselines}. With an additional CNN, {\tool-CM} can increase up to \textbf{2.75\%} over basic LSTM model.
\end{tcolorbox}

\subsection{Performance of \tool-CR (RQ2)}
\label{ss-rq2}

Similarly, we evaluate our \tool-CR with a $K$-fold stacking algorithm, $K$fs (code), and basic LSTM model, LSTM (code). Furthermore, we added one evaluation against a statement-level LSTM without CNN, LSTM (LineReps). Table~\ref{tbl-result} shows the results on code revisions for each dataset. 

\tool-CR outperforms the $K$fs (code) for FFmpeg, Qemu and the combined dataset, in terms of F1-Score, by \textbf{2.72\%}, \textbf{6.05\%} and \textbf{1.54\%} respectively. Furthermore, \tool-CR also outperforms both LSTM (code) and LSTM (LineReps), in terms of F1-Score, by at least \textbf{11.3\%} in these three projects. 

However, we observed that there are two exceptions to this evaluation. Firstly, $K$fs (code) in Linux outperforms all other models. This aligns with our reasoning in the previous RQ as the data of the Linux dataset are imbalanced and, hence, $K$fs (code) could predict more positive cases than negative, resulting in a higher F1-score. Secondly, for Wireshark, LSTM (code) outperforms \tool-CR by \textbf{1.36\%} in terms of F1-score. One of the possible reasons is that Wireshark has a higher average line of code at 29.52. This resulted in statement-level models do not perform as good as expected. 

\emph{Note.} Compared to the results on code revisions, the results on commit messages are better. This is mainly because code revisions have longer sequences and more complicated structure and logic. We take the first maximum length of tokens in code revisions for learning and drop the rest. It may cause us to miss some representative code fragments. If we take a longer sequence, however, it will go beyond the capacity and capability of contemporary neural networks, which causes inferior performance in the other way.

\begin{tcolorbox}[breakable,width=\linewidth,boxrule=0pt,top=1pt, bottom=1pt, left=1pt,right=1pt, colback=gray!20,colframe=gray!20, ]
\textbf{Answer to RQ2:} Our proposed neural network \tool-CR for learning on code revisions learns useful representations for detecting security patches. The performance is not as good as \tool-CM, but outperforms $K$-fold stacking method and basic LSTM models.
\end{tcolorbox}
\vspace{-2mm}
\subsection{Performance of {\tool} (RQ3)}
\label{ss-rq3}

We scrutinize the combined performance of the learned results from \tool-CM and \tool-CR. Table~\ref{tbl-combined-result} shows the scores of \tool. We observed that the approach of weighted combination provides a great improvement in the F1-score, as high as \textbf{11.18\%} for the Qemu dataset, as compared to \tool-CM and \tool-CR. The trained model on the combined dataset achieved \textbf{89.68\%} in recall as well as precision of \textbf{86.24\%}, and the highest F1-score of \textbf{87.93\%}. This shows that the combination of features in code revision and commit messages could help identify security patches better.

\begin{tcolorbox}[breakable,width=\linewidth,boxrule=0pt,top=1pt, bottom=1pt, left=1pt,right=1pt, colback=gray!20,colframe=gray!20]
\textbf{Answer to RQ3:} Despite the lacking performance in \tool-CR, {\tool} \jingkai{achieves better results} in combining the features in commit messages and code revision. With an increment, as high as 11.18\%, in F1-score, {\tool} can effectively identify undisclosed security patches based on the commit message and code revision.
\end{tcolorbox}

\subsection{Performance of {\tool} in Cross-Project Evaluation (RQ4)}
\label{ss-rq4}

\jingkaicyan{\emph{SPI} can effectively identify security patches when the training data and testing data are from the same open-source projects, as shown in the above RQs and experiments. We further evaluate the generalization ability of our approach by cross-project evaluation, \emph{i.e.,} evaluating whether \emph{SPI} can identify security patches from projects that are different from the training dataset. Table~{\ref{tbl-cross-result}} shows the result of the cross-project evaluation. In each experiment, we isolate one of the projects as a testing dataset and used the remaining three projects as the training dataset. For example, in LinuxTest, we used Linux as the testing dataset, while we trained our model on FFmpeg, Qemu, and Wireshark. }

SPI performs well under the setting of cross-project evaluation, achieving an F1-Score of 82.45\%, 87.33\%, 73.01\%, and 81.96\% for LinuxTest, FFMpegTest, QemuTest, and WiresharkTest respectively. On average, SPI performs better than \emph{K}fs(msg) by 8.13\% and \emph{K}fs(code) by 30.11\%. Comparing to LSTM(msg) and LSTM(LineReps), F1-Score increased, on average, by 1.78\% and 15.80\% across the four experiments. We observed that, despite a decrease in performance as compared to inter-project evaluation, \emph{SPI}-CM achieves F1-score between 72.95-87.18\% and \emph{SPI}-CR achieves an F1-score in a range of 60.63-82.13\%, while \emph{SPI} achieves F1-Score between 73.01-87.33\%. As compared to RQ3, the performance of \emph{SPI}-CM and \emph{SPI}-CR across four experiments decreases by 7.03\% and increases by 7.32\% on average. However, the overall performance of \emph{SPI} decreases by 11.14\% across all four experiments.

\jingkaicyan{The decline in the performance of \emph{SPI} is inevitable since we evaluate the model using different projects. However, \emph{SPI} still can perform well, in the range of 73.01-86.96\% in F1-Score. This demonstrates the ability to identify security patches that are from different domains despite the absence in the training dataset. Hence, our approach can identify cross-domain security patches even with the lack of the project data in the training set.}

\begin{tcolorbox}[breakable,width=\linewidth,boxrule=0pt,top=1pt, bottom=1pt, left=1pt,right=1pt, colback=gray!20,colframe=gray!20]
\textbf{Answer to RQ4:} \jingkaicyan{Wide range of projects should be incorporated into the projects to allows the performance of the \emph{SPI} to generalize even better to unseen security patches. Our model is shown to generalize well to unseen projects and can identify security patches from projects that are not in the dataset. } 
\end{tcolorbox}

\subsection{Production Observation with {\tool} (RQ5)}
\label{ss-rq5}

We discuss the effectiveness and practicality of our model in a production setting. We deployed our pipeline in a production test of our industry collaborator. We crawled a different set of data from 410 C Language open-source libraries. The production dataset consists of 298,917 commits after the filtering process. As the data come from various C projects, they exhibit distinct data distributions and features, such as variable names that are unique to projects or longer source code. 

\textbf{Production Prediction.}We used the same model that was trained using the combined dataset to predict all the commits in the production dataset. We employed the same setting, \emph{i.e} word embedding model, neural networks structure, and hyperparameters, as in the testing phase (Section \ref{sec-eval}). Out of 298,917 commits in the production dataset, 136,466 (45.65\%) were predicted as SPs and 162,451 (54.34\%) as NSPs. 

\textbf{Prediction Verification.} To ensure that our model works properly, we manually verified 1146 commits that were picked randomly from the predicted NSPs and found out that \textbf{93.63\%} were predicted correctly. The high precision on predicted NSPs implies that non-security patches can be effectively filtered out from an unlabeled dataset. In other words, we could effectively confirm \textbf{151,080} out of the 162,451 predicted NSPs. This reduces the workload and time needed for the manual verification process.  
By the time of this paper, 43,543 out of the 136,466 predicted SPs had been verified manually, with an initial precision of \textbf{61.49\%}. \jingkaiorange{We further manually verified 1000 random commits from the industrial dataset and observed that there are 384 security patches among these commits, \emph{i.e.} 38.4\% of the 100 verified commits are security patches. As shown in Table 4, the four selected projects have security patches in ranges of 41\% to 69\%, the sampling from the industrial dataset shows that security patches are even more scarce in the wild, highlighting that our work is important. The gap in proportion between the industrial dataset and our four selected projects dataset is caused by the diversity of the industrial datasets, which includes 410 C Language open-source libraries while the experimental datasets consist of only 4 libraries. Hence, the F1-Score is lower than the test results, which possibly caused by the distribution of SP ratio and the diversity of the two datasets.}


\textbf{Iterative Model Training}. It is challenging to get a sizable and high-quality \emph{labelled} dataset at one time to train well-generalized models. In our case, the initial trained model has helped significantly accelerate manual validation of new data, where the labelled new data can be fed back into the learning process. By iterating over training and manual validation, it forms a closed-loop for product development and improves performance gradually. Overall, the initial production test proves the usability and effectiveness of our system. We iterated and retrained the model with a dataset that is a combination of training dataset and production dataset. We then obtained an improved model with a precision of \textbf{77.99\%}. This shows that the model can be improved iteratively in a production setting.

\begin{tcolorbox}[breakable,width=\linewidth,boxrule=0pt,top=1pt, bottom=1pt, left=1pt,right=1pt, colback=gray!20,colframe=gray!20]
\textbf{Answer to RQ5:} The deployment of {\tool} ~in production validated its usability and effectiveness. \jingkai{Instead of manually verifying 298,917 commits, {\tool} reduce the amount of verifying commits by 50.54\% by fulfilling as a pre-processing step. It saves manual workload effectively in building security patch database, and drives to build more robust and general prediction models with data and product iteration. }

\end{tcolorbox}

\subsection{Hyper-Parameter Tuning with {\tool-CM}, {\tool-CR}, and {\tool} (RQ6)}
\label{ss-rq6}

\begin{figure}[tp]
	\centering
	\includegraphics[width=1.0\textwidth]{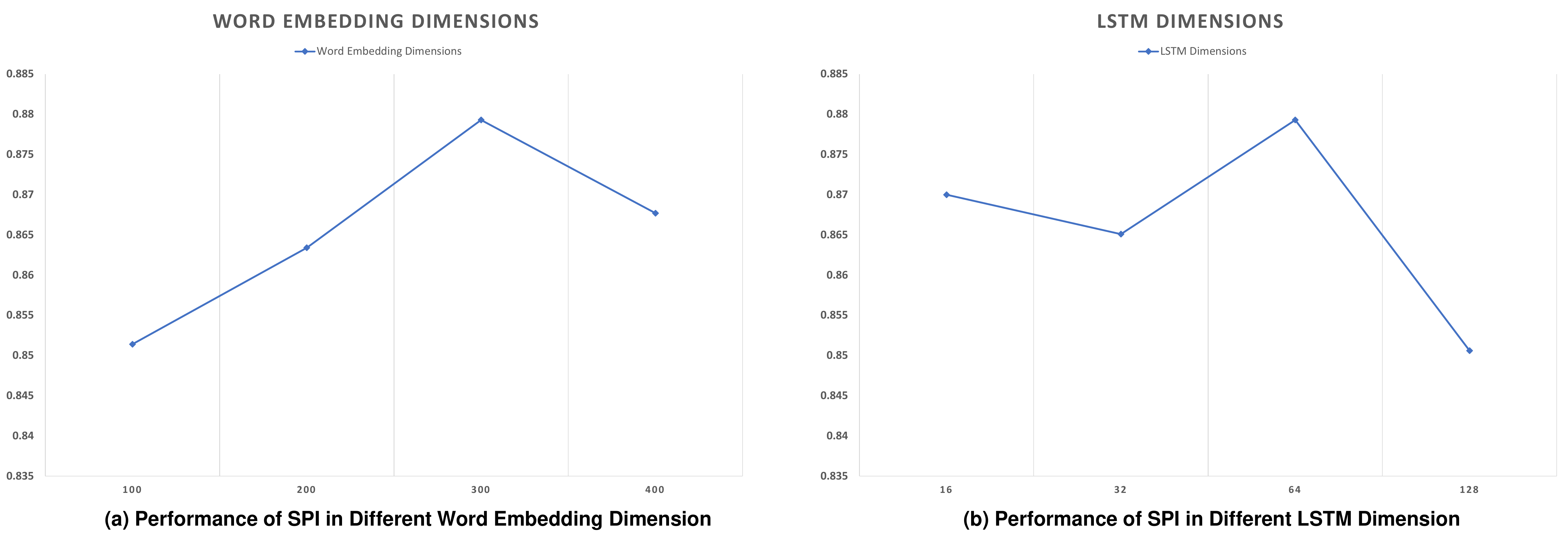}
	\caption{Hyper-Parameter Tuning}
	\label{fig-hyper-parameter-tuning}
\end{figure}

\jingkai{We explore two hyper-parameters tuning, word embedding dimensions, and the dimension of LSTM, in RQ6. In our work, we employ word2vec{~\cite{mikolov2013distributed}} as our method for embedding textual data into their numerical representation. The length of the vector representation can affect the performance of the deep learning model. If the length of the vector representation is too large, the training of the model might slow down, failing to converge properly. On the other hand, if the length of the vector representation is too small, the information of the textual data cannot be an effective representation. }

\jingkai{The dimensions of LSTM refer to the number of learned parameters in each cell. Similarly, it affects the learning capability of the model. If the dimension of LSTM is too small or too large, the performance of {\tool} might not be optimal. Therefore, in this RQ, we experiment on different values of these hyper-parameters. We further show that the parameters that are used in the previous RQs allow us to achieve the best performance in {\tool}.}

\jingkai{Figure{~\ref{fig-hyper-parameter-tuning}} shows the performance of {\tool} in different parameter settings. In Figure{~\ref{fig-hyper-parameter-tuning}}(a), we experiments with different word embedding dimensions, ranging from 100 to 400.}

\jingkai{As observed from the table, the performance of {\tool} increases proportionately as the dimensions of the word embedding increases. However, it decreases slightly after the dimension reaches 300. Figure{~\ref{fig-hyper-parameter-tuning}}(b) shows the tuning of LSTM units. We observed from our experiments that the performance of {\tool} shows an upward trend despite that the trend is non-monotonic. However, we discovered that our performance of {\tool} peaks when we select the value of the LSTM unit to be 64. }

\begin{tcolorbox}[breakable,width=\linewidth,boxrule=0pt,top=1pt, bottom=1pt, left=1pt,right=1pt, colback=gray!20,colframe=gray!20]
\textbf{Answer to RQ6:}  \jingkai{Deep learning networks are sensitive to hyper-parameters as they affect the learning capability of the networks. We showed hyper-parameters tuning on two main hyper-parameters, word embedding dimension and LSTM dimension, for exploration of the best values. Our experiments show that the performance of {\tool} peaks when the word embedding dimension is 300 and LSTM unit is 64. }

\end{tcolorbox}

\section{Discussion}
\label{sec-dis}

\subsection{Impact of Sequence Length on \tool}
\begin{table}[t]
	\centering
	\caption{Performance of {\tool} for different message length}
	\label{tbl-msg-length}
	\begin{tabular}{|l c c c c |}
		\hline 
		\textbf{Msg Length ($l$)} & \textbf{Total} & \textbf{SP} & \textbf{NSP} & \textbf{F1} \\
		\hline 
		\textbf{$l<$50} & 5609 & 2558 & 3051 & \textbf{0.9217} \\
		\hline 
		\textbf{50$\leq l<$100} & 2593 & 1385 & 1208 & 0.8435 \\
		\hline 
		\textbf{100$\leq l<$155} & 936 & 520 & 416 & 0.8267 \\
		\hline 
		\textbf{$l>$150} & 993 & 635 & 358 & 0.8085 \\
		\hline 
	\end{tabular}
	\vspace{-3mm}
\end{table}

\begin{table}[t]
	\centering
	\caption{Performance of {\tool} for different code length}
	\label{tbl-code-length}
	\begin{tabular}{|l c c c c |}
		\hline 
		\textbf{Code Length ($l$)} & \textbf{Total} & \textbf{SP} & \textbf{NSP} & \textbf{F1} \\
		\hline 
		\textbf{$l<$50} & 3643 & 2052 & 1591 & \textbf{0.8976} \\
		\hline 
		\textbf{50$\leq l<$100} & 1995 & 1069 & 926 & 0.8943 \\
		\hline 
		\textbf{100$\leq l<$155} & 1072 & 527 & 545 & 0.8761 \\
		\hline 
		\textbf{$l>$150} & 3421 & 1450 & 1971 & 0.8315 \\
		\hline 
	\end{tabular}
	\vspace{-3mm}
\end{table}

We investigated the performance of {\tool} on different commit messages and code-revision length. We observed that for both code-revision and commit message, {\tool} predicted more accurately for sequences with length less than 50. Table \ref{tbl-code-length} and Table \ref{tbl-msg-length} shows the results of our experiments. {\tool} can predict more accurately on message with length less than 50 with an f1-score of \textbf{92.17\%} and on code with length less than with an f1-score of \textbf{89.76\%}.  One of the possible reasons for better performance among shorter code revision and commit messages might be that security patches are often small and quick fixes to the codebase. Commonly, exploits and loopholes are patched with additional checkings, such as sanity checks. Longer code-revisions are hard to learn as they are complex and often span across different parts of the file. 

\subsection{Prediction on Implicit Security Patches} \label{implicit}
We are motivated to investigate the effectiveness of learning from code revisions alone on finding security patches. \jingkai{In our paper, we define implicit security patches as patches that fix a vulnerability but their commit messages does not convey their intention of fixing the vulnerability.} To avoid disputation, we use the fix commits of CVEs that are officially verified by vendors and security professionals.
We use the following four CVEs and their security patches as an evaluation of our model.


\textbf{CVE-2017-7187} \cite{CVE1}. This is a high-risk Denial of Service (DoS) from Linux Kernel, caused by a large-size command from user data. 
The commit message and code revision is shown in Table~\ref{tbl-cve-code-example}.
Despite the message mentioning the checking the length of SG\_NEXT\_\~CMD\_LEN because a user can control the size of the command, it is unclear from the message if it is a patch. By looking at the code revision, it returned a token, -ENOMEM, if the input size is larger than the allowed size, indicating that before the fix the user can manipulate much larger size data in memory. 

\textbf{CVE-2017-8063} \cite{CVE2}. This high-risk DoS vulnerability was disclosed in the Openwall community \cite{openwall} by pointing out that it actually was silently fixed by the Linux Kernel developers without mention of it at all in the commit message.

Similar examples can be found in CVE-2010-5329 \cite{CVE3}, CVE-2015-8952 \cite{CVE4}, and etc, where commit messages of the patches do not mention they were meant to fix vulnerabilities. Table \ref{tbl-cve-code-example} shows the predict results, commit message and change revisions of CVE-2017-7187. The commit message has no security-related words shown in Table~\ref{tbl-keywords} or indication of fixing any vulnerability, thus \tool-CM predicted it as negative. On the other hand, as we analyzed previously, the code revision could be an indication for fixing a vulnerability, which is predicted as a security patch by \tool-CR. The combined result by {\tool} is positive as the final prediction. Out of the four CVE patches, we achieve about 75\% accuracy, where patches of CVE-2017-7187, CVE-2017-8063 and CVE-2010-5329 are predicted correctly as security patches though the message predictor failed. Only CVE-2015-8952 was predicted wrongly because of the long code revision in the patch.

As seen from these implicit security patches, our code-revision model provides meaningful insights to decide whether the commit is a security patch. The commit messages might not be able to provide sufficient information in deducing the type of commits. The code-revision model allows us to draw insights on the commits based on the code-revision. Therefore, the evaluation proves that the code-revisions models are essential in identifying security patches.

\subsection{Prediction on Explicit Security Patches}

We show an example of a security patch, where \tool-CR correctly predicts it as a security patch based on its commit message while \tool-CM fails. Our example is a commit \cite{commitfm} from FFmpeg. We perform predictions using {\tool-CR}, {\tool-CM} and {\tool} to evaluate the different component of our approaches. Table \ref{tbl-cve-msg-example} shows the commit message and prediction results. The code-revision is omitted from the table due to the massive revision. Please refer to \cite{commitfm} for the code-revision. The shown example has a short message with evidence of fixing some buffer overreads. Buffer overread \cite{buffover} is a common, yet serious, vulnerability that could lead to more severe consequences, such as privilege escalation and denial of service. 

On the other hand, the code-revisions are long and spans over several different parts of the functions. It consists of changes of several function calls and complex equations. This prevents the neural networks to learn that it is a security patch. The prediction results of {\tool} show that our model can also effectively identify such security patches without any additional information.

\begin{table}
	\begin{center}
		\caption{Evaluation on FFmpeg commit}
		\label{tbl-cve-msg-example}
		\begin{tabular}{p{2cm}|p{5.9cm}}
		    \hline
		    Commit ID & 61cd19b8bc32185c8caf64d89d1b0909877a0707 \\
			\hline
			Message & vmnc: Port to bytestream2 Fix some buffer overreads. \\
			\hline
			\tool-CM & 1 \\
			\hline
			\tool-CR & 0 \\
			\hline
			\tool & 1 \\
			\hline
		\end{tabular}
	\end{center}
\end{table}

\begin{table}
	\begin{center}
		\caption{Evaluation on the fix commit of CVE-2017-7187}
		\label{tbl-cve-code-example}
		\begin{tabular}{p{2cm}|p{5.9cm}}
			\hline
			Message &
			scsi: sg: check length passed to SG\_NEXT\_CMD\_LEN
            The user can control the size of the next command passed along, but the
            value passed to the ioctl isn't checked against the usable max command
            size.  \\
			\hline
			Code revision & \Plus\emph{ 	if $($val $>$ $SG\_MAX\_CDB\_SIZE)$ \newline
+			return $-ENOMEM;$  } \\
			\hline
			\tool-CM & 0 \\
			\hline
			\tool-CR & 1 \\
			\hline
			\tool & 1 \\
			\hline
		\end{tabular}
	\end{center}
\end{table}

\section{Threats to Validity}
\label{sec-dis}
\noindent\textbf{Data Quality.} We understand that the judgment of vulnerabilities is subjective to the human expert's expertise and their experience. Therefore, our security researchers that involved in the manual labelling process are experienced enough to be able to determine security patch. Thus we believe that our manual labels are fair. 
Our datasets have a lack of implicit security patch samples and also contain false negatives by human errors due to manual labelling. The resulted model may overlook this kind of security patch because of the bias in the training data. To address it, 1) we can collect more correctly labelled data samples to make the learning on messages more robust, 2) we can put more weight on the predicted result by code revisions for this kind of commits as we have shown that security patches with implicit commit messages can be found through the code revision network.

\noindent\textbf{Evaluation Metrics.} We believe our chosen metrics are sufficient in evaluating our model and approaches as F1-score are a common evaluation metric in other similar works for binary classification \cite{8115623, fse2017vul}. Furthermore, our dataset is fairly balanced, having 50.59\% of positive label and 49.41\% of the negative label. Hence, F1-score is suitable for our approach. 

\noindent\textbf{Language Generalization.} It is important to generalize the proposed approaches to projects of other programming languages. Despite the keywords that differ from projects to projects, the approach of extracting keywords remains the same. We believe that by collecting high-quality datasets with a larger size, it is promising to apply our proposed system to effectively detect patches in other languages. As opposed to handcrafted features, our approach could apply to other languages as long as the word embedding is trained using a corpus of the chosen language. 

\noindent\textbf{Deep Learning on Code Revision.}
The overall prediction results on commit messages are better than code revisions.
1) Most of our data are collected and labelled mainly by human comprehension on commit messages. Hence, the characteristic in commit messages is more dominating and easier to learn than code revision. Thus the test result on commit messages is too high to improve in the collected data sets. 2) We found that if the changes between negative and positive code revision are minor, the neural networks fail to pick up the changes and cause poor performance. We conduct a similarity analysis on the code tokens in both subtractive revision and additive revision in the training data. For each commit, we compute the percentage of matching words in both revisions. About 40\% of the commits have less than 10\% of similarity and about 10\% of the commits have more than 60\% of similarity percentage. This is one of the reasons why our code-revision model does not perform as well as the commit message model. 3) Code revisions have complicated structures where natural sequencing of code tokens cannot represent them sufficiently. More complex representations, such as graph structure, can be incorporated into our approach. We will explore encoding the rich semantic information of these classical code property graphs into graph neural networks \cite{li2015gated}, and learn comprehensive programming semantics for better performance.

\section{conclusion}
\label{sec-con}
We have initiated the study of the deep neural-network-based approach to automatically classify security patches, built vulnerability-commit datasets, and integrated them into the industrial production cycle. The prediction system effectively learns from commit messages and code revisions and consolidates the learned results together to reach conclusive decisions. Our work demonstrates that it is promising to apply deep neural networks to scale up patches identification via an automatic and evolutionary approach and improve the state of the art \cite{fse2017vul} in the industry.

\section{Acknowledgement}
This research is supported by the National Research Foundation, Singapore under its AI Singapore Programme (AISG Award No: AISG2-RP-2020-019), the National Research Foundation under its National Cybersecurity R\&D Program (Award No. NRF2018NCR-NCR005-0001), the Singapore National Research Foundation under NCR Award Number NRF2018NCR-NSOE003-0001 and NRF Investigatorship NRFI06-2020-0022. We are grateful for the GPU support of Nvidia.

\newpage
\bibliographystyle{ACM-Reference-Format}
\bibliography{ref}
 
\end{document}